\documentstyle[amsmath,amssymb,12pt,epsf]{article}
\begin{document}
\begin{flushright} 
{CU-TP-1060}
\end{flushright}
\vskip50pt
\begin{center}
\begin{title}
\title{\large\bf Gluon distributions and color charge correlations\\
in a saturated light-cone wavefunction\footnote{This research is supported in part by the US Department of
Energy.}}\\

\vskip 10pt
{A.H. Mueller\\

Physics Department, Columbia University\\
New York, N.Y. 10027 USA}
\end{title}
\end{center}
\vskip 10pt
\noindent
{\bf Abstract}

We describe the light-cone wavefunction in the saturation regime in terms of the density of gluons per unit of transverse phase space, the occupation number, and in terms of the color charge correlator.  The simple McLerran-Venugopalan model gives what are claimed to be general results for the  phase space  gluon density, but it does not well describe the general case for the charge  correlator.  We derive the general momentum dependence and rapidity dependence of the color charge correlator which exhibits strong color shielding.  A simple physical picture which leads to these general results  is described.

\section{Introduction}
Our object in this paper is to describe what the light-cone wavefunction looks like in the saturation regime, that is when the occupation numbers of gluons are large.  What we attempt to describe is the density of gluons and the color charge correlator as a function of rapidity and transverse momentum.  We shall carry out this description first for the McLerran-Venugopalan model\cite{ran,ian,gov}, then for an unsaturated BFKL\cite{Kur,sky} light-cone wavefunction and finally for a more general saturated wavefunction, both for fixed coupling and for running coupling.

For the gluon density in phase space we find the result that ${dxG\over d^2bd^2\ell_\perp}$ is proportional to ${N_c^2-1\over \alpha N_c} \ell n[Q_s^2(Y)/\ell_\perp^2]$ in all the saturation models. We believe this to be a very general result which follows straightforwardly from the physical picture we give in Sec.6.

The color charge correlator $\rho(\ell_\perp,Y),$ defined below in (17), is a constant in the McLerran-Venugopalan model reflecting the fact that the elementary charges are completely uncorrelated in that model on scales small compared to one fermi.  We note that the first gluonic corrections to the  model, calculated in Sec.2 begin to show charge shielding.  Charge shielding is clearly visible in the BFKL wavefunction and in saturated wavefunctions produced from BFKL evolution.  In the latter case, both for fixed coupling and for running coupling, ${d\over dY}\rho(\ell_\perp,Y) = {N_c^2-1\over \pi} \ell_\perp^2$ in the saturation region, and the proportionality to $\ell_\perp^2$ means that charge shielding is as complete as it can be.  That is when measured on a scale $\Delta x_\perp \sim 1/\ell_\perp$ all gluons having momenta $k_\perp\gg \ell_\perp$ contribute not at all to the charge correlator.

In Sec.6 we present a simple picture which leads to the results quoted for${dxG\over d^2bd^2\ell}$ and for ${d\over dY} \rho(\ell_\perp,Y).$  This picture is most easily explained by referring to Figs. 5 and 6 where the saturation region is shown.  In Fig.5 the saturation region is illustrated for fixed coupling in the $y$ and $\ell n\  k_\perp^2/\Lambda^2$ plane.  Note that $y$ increases in the \underline{downward} direction in the figure.  The diagonal line is the separation line between the dilute and dense regimes.  In Fig.6 the saturation region is shown as a function of $x_-$ and $y$ in the Kovchegov gauge\cite{Kov}.  (This picture is that first given by Iancu, Leonidov and McLerran\cite{Ian} where a gauge was used which puts all the gluons in saturation at positive values  of $x_-,$ rather than the choice here of negative values of $x_-$\cite{Kov,ler}.) In terms of these  figures, gluons contributing to ${d\over dY} \rho(\ell_\perp,Y)$ are located in the shaded region \ C\ and they come by direct bremsstrahlung emission from the shaded region A.  In the region $A, \ell_\perp$ is the saturation momentum of that rapidity and so the sources at A  do not have strong shielding on the scale $\ell_\perp$ and thus the gluons they emit into region \ C,\ having a comparable transverse momentum $\ell_\perp,$ are also not far from random in color.  This naturally gives ${d\over dY} \rho(\ell_\perp,Y) \propto \ell_\perp^2$ with the $1/\alpha$ density of sources being cancelled by the $\alpha$ for a gluon emission leaving ${d\over dY} \rho$ independent of $\alpha.$  Gluons can come into the region \ C\ from regions other than A  but they do not contribute to ${d\over dY} \rho$ because they are strongly shielded, as explained in Sec.6.
 
Now turn to ${dxG\over d^2bd^2\ell},$ evaluated at $Y.$  Such gluons come generically from anywhere along the saturation line in Fig.5, the region  \ B\ being typical. In Sec.6 we show how emissions from the saturation line, between $Y(\ell_\perp)$ and $Y,$  naturally lead to ${dxG\over d^2\ell_\perp d^2b} \sim {1\over \alpha} \cdot \ell n[Q_s^2(Y)/\ell_\perp^2]$\cite{ian,ler,Mue,ler}.

Finally, in Sec. 6 we show how the $x_--$ distribution of saturated gluons, as shown in Fig.6, naturally leads to a Glauber-like picture of the scattering of a color dipole on a saturated wavefunction.  In passing the saturated hadron the dipole encounters, in a sequential manner, gluons located at different rapidities and scatters independently off them thus naturally leading to the exponential result required by the Balitsky-Kovchegov equation\cite{Bal,Phy,vin}.

\section{The McLerran-Venugopalan model[1-3]}
\subsection{The valence quark approximation}

In the leading approximation the wavefunction of a high momentum large nucleus consists of the valence quarks of the various nucleons of the nucleus Lorentz-contracted to a thin disc.  So long as transverse distances less than a fermi are considered the quarks may be viewed as uncorrelated in color.  At an impact parameter \ b\  from the center of the nucleus the number of valence quarks per unit area is

\begin{equation}
N(b) = N_0 N_c 2{\sqrt{R^2-b^2}}
\end{equation}

\noindent where $N_0$ is the nuclear density, the number of nucleons per unit volume in the rest system of the nucleus, $N_c$ is the number of colors, and \ R\ is the radius of the  nucleus.  Throughout this paper we take a simplistic view of a nucleus as having a uniform density, $N_0,$ of nucleons within a sphere of radius  R.

Suppose we take the nucleus to be moving along the positive z-axis.  Now scatter a left-moving quark-antiquark dipole, moving along the negative z-axis, on the nucleus.  The probability that the dipole pass through the nucleus without interacting is equal to

\begin{equation}
S^2(b, x_\perp) = e^{-N(b)\sigma_q(x_\perp)}
\end{equation}

\noindent where $x_\perp$ is the transverse separation of the quark-antiquark pair of the dipole and \ S\ is the \ S-matrix for elastic scattering of the dipole with the nucleus.  $\sigma_q$ is the cross-section for scattering of the dipole on a single quark which we take to be\cite{tel} 

\begin{equation}
\sigma_q(x_\perp) = {\pi^2\alpha\over N_c} x_\perp^2xG_q(x, 1/x_\perp^2)
\end{equation}

\noindent where the gluon distribution in a quark is given as

\begin{equation}
x G_q(x, 1/x_\perp^2) = {\alpha C_F\over \pi} \ell n\left({r_0^2\over x_\perp^2}\right)
\end{equation}

\noindent in the two-gluon exchange approximation illustrated in Fig.1.  In (4)  $r_0$ is the radius of a nucleon and we always suppose that $x_\perp^2/r_0^2 \ll 1.$  The S-matrix can be written as

\begin{center}
\begin{figure}
\epsfbox[0 0 288 57]{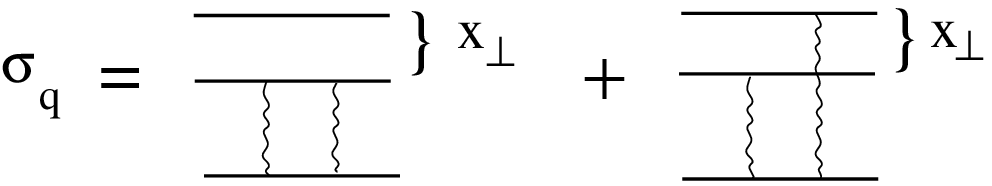}
\caption{}
\end{figure}
\end{center}

\begin{equation}
S = e^{-x_\perp^2{\bar{Q}}_s^2/4}
\end{equation}

\noindent where the saturation momentum measured by quarks is given as

\begin{equation}
{\bar{Q}}_s^2=C_F/N_c Q_s^2
\end{equation}

\noindent and where\cite{Mue} the gluon saturation momentum is

\begin{equation}
Q_s^2 = {4\pi^2\alpha\over N_c^2-1} N(b) x G(x, Q_s^2)
\end{equation}

\noindent with the gluon distribution of a nucleon $x G$ given by

\begin{equation}
x G((x,Q_s^2) = N_c x G_q(x, Q_s^2).
\end{equation}

Finally, the color charge correlator in the nucleus takes the form\cite{dov}

\begin{equation}
<\rho^i(b+x_\perp)\rho^j(b)>_{MV}={\delta_{ij}\over N_c^2-1} \delta^2(x_\perp) g^2C_FN(b)
\end{equation} 

\noindent where we use a normalization such that

\begin{equation}
<\rho^i(b+ x_\perp) \rho^j(b)>_{quark}=\delta_{ij}\delta^2(x_\perp){g^2\over 2N_c}
\end{equation}

\noindent for the color correlator of a single high momentum quark.

\subsection{Valence quarks along with Weizs\"acker-Williams gluons}

The valence quarks of a fast nucleus can emit (virtual) gluons which then form the gluon distribution of the nucleus in the semiclassical, Weizs\"acker-Williams, approximation.  The well-known result for the distribution of gluons is\cite{ian,ler}

\begin{equation}
{d xG_A\over d^2\ell_\perp d^2b} = \int{d^2x_\perp\over 4\pi^4}\ {N_c^2-1\over \alpha N_cx_\perp^2} (1-e^{x_\perp^2Q_s^2/4})e^{i\ln_\perp\cdot x_\perp}
\end{equation}

\noindent or

\begin{equation}
{d x G_A\over d^2\ell_\perp d^2b}\  \ {_{_\approx}\atop \ell_\perp^2/Q_s^2\ll 1}\ \ {N_c^2-1\over 4\pi^3\alpha N_c}\  \ell(Q_s^2/\ln_\perp^2).
\end{equation}

\noindent ${(2\pi)^3\over 2(N_c^2-1)}\ {d x G_A\over d^2\ell_\perp d^2b}$ is the occupation number of gluons, per unit of rapidity.  We believe (12), except for overall normalization, is a general result for the phase space density of gluons deep in the saturation regime, $\ell_\perp^2/Q_s^2\ll1,$ of a high energy wavefunction\cite{Mue,Ler}.

Now consider the scattering of a dipole on the nucleus, including the semi-classical gluons in the wavefunction of the nucleus.  We wish to determine how (5) and (9) change in the presence of the Weisz\"acker-Williams gluons in the nucleus.  The calculation is most easily done by Lorentz transforming the dipole-nucleus system so that the dipole carries a greater momentum and the nucleus a lesser momentum.  In this way we may view the Weisz\"acker-Williams gluons involved in the interaction as being part of the wavefunction of the dipole.  The scattering which modifies (5) now can be taken to be that of a left-moving quark-antiquark-gluon system on a McLerran-Venugopalan nucleus having only valence quarks in its wavefunction.  In the large $N_c$ limit this contribution, $\delta S,$ is given by

$$\delta S(b,x_\perp) = \Delta Y{\alpha N_c\over 2\pi^2}\cdot $$

\begin{equation}
\cdot \int {d^2z_\perp x_\perp^2\over (x_{1\perp}-z_\perp)^2(z_\perp-x_{2\perp})^2} S(b, x_{1\perp}-z_\perp) S(b, z_\perp-x_{2\perp})
\end{equation} 

\noindent and is illustrated in Fig.2.  Deep in the saturation regime, ${\bar{Q}}_s^2 x_\perp^2\gg 1,$ the dominant contribution to (13) comes when $z_\perp = {x_{1\perp} + x_{2\perp}\over 2}.$  One finds

\begin{center}
\begin{figure}
\epsfbox[0 0 187 139]{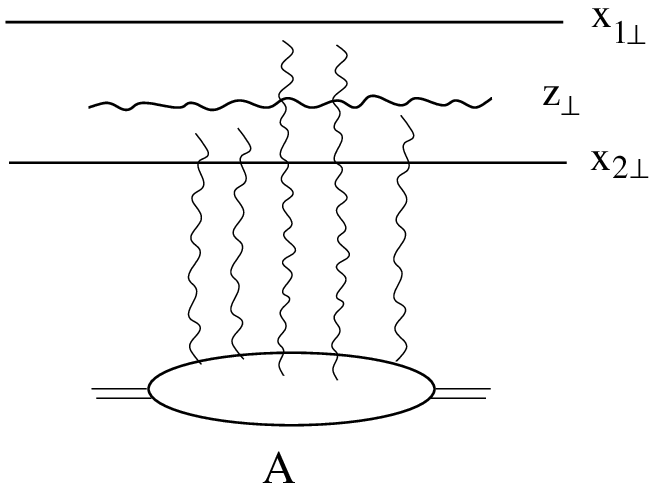}
\caption{}
\end{figure}
\end{center}

\begin{equation}
\delta S(b,x_\perp) = {16\alpha N_c\over \pi}\ {\Delta Y\over {\bar{Q}}_s^2x_\perp^2}\ e^{-x_\perp^2{\bar{Q}}_s^2/8}
\end{equation}

\noindent where $\Delta Y$ is the range of rapidity of the gluons we are considering. Eq.14 shows that the transmission amplitude actually grows and becomes greater for a quark-antiquark-gluon to pass through the nucleus without interaction than for a quark-antiquark to pass through without interaction.  In the frame in which we have done the calculation the extra gluon tends to shield the charges in the dipole.

In a frame where the Weisz\"acker-Williams gluons are in the nuclear wavefunction, and (14) is given by the gluon dependent part of the scattering of a quark-antiquark dipole on the nuclear wavefunction, the shielding is expressed by

\begin{equation}
<\rho^i(x_\perp + b) \rho^j(b)>_{WW} = {1\over 2} <\rho^i(x_\perp + b) \rho^j(b) >_{MV}.
\end{equation}

\noindent This shielding continues to grow as more and  more gluons are created.  The McLerran-Venugopalan valence quark model seems to be a good starting point for calculating the density of gluons in the saturation region, as given by (11) and (12), but it is not a good starting point for evaluating the distribution of color charge in a large nucleus.  In the McLerran-Venugopalan model charges are randomly distributed while in the actual wavefunction there is an arrangement of the quarks and gluons so as to give strong color shielding. Eq.(15), or (14), is the first indication of that shielding.

\section{The BFKL wavefunction}

The BFKL\cite{Kur,sky} wavefunction in a dilute (non-saturation) regime already exhibits the charge shielding which we shall find in the saturation regime at extremely high energies. The gluon distribution in the BFKL wavefunction of a high-energy hadron is related to the charge correlator of gluons {\underline {at all higher rapidities}} by

\begin{equation}
{dxG\over d^2\ell_\perp} = {1\over 4\pi^3\ell_\perp^2} \rho(\ell_\perp,Y).
\end{equation}

\noindent Our notation is such that

\begin{equation}
\rho(\ell_\perp,Y) = \int d^2x_\perp e^{-i\ell_\perp\cdot x_\perp} \sum_i <\rho^i(b + x_\perp) \rho^i(b) > d^2b
\end{equation}

\noindent and where the $Y$  variable now indicates a dependence on rapidity in contrast to the McLerran-Venugopalan model.  Using

\begin{equation}
{dxG\over d^2\ell_\perp} \ = c\ {e^{(\alpha_P-1)Y}\over {\sqrt{\alpha Y}}}\ {1\over \ell_\perp}
\end{equation}

\noindent for fixed coupling BFKL evolution one finds

\begin{equation}
\rho(\ell_\perp,Y) = 4\pi^3c \ell_\perp\ {e^{(\alpha_P-1)Y}\over {\sqrt{\alpha Y}}}.
\end{equation}

\noindent The corresponding result using (9) is

\begin{equation}
\rho_{MV}(\ell_\perp) = g^2C_F\cdot N_q
\end{equation}

\noindent with $N_q$ the total number of quarks in the nucleus.  Aside from the Y-dependence in (19), which one expects will disappear in the saturation regime, the big difference between (19) and (20) is that $\rho_{MV}$ has no dependence on $\ell_\perp$ while the BFKL correlator grows linearly with $\ell_\perp.$  The lack of an $\ell_\perp-$dependence in $\rho_{MV}$ is natural and expresses the fact that the quarks are completely uncorrelated in color in our short distance approximation.  The growth of the BFKL correlator with\ $\ell_\perp,$ as given by (19), indicates strong color shielding.  The linear growth in $\ell_\perp$ in (19) means that source gluons having $k_\perp \ll \ell_\perp$ contribute little to $\rho(\ell_\perp,Y)$ since there are two few such gluons while source gluons having $k_\perp\gg \ell_\perp$ contribute little to $\rho(\ell_\perp,Y)$ so that such high momentum gluons are perfectly shielded.  Consider an ``area'' in momentum space $\Delta \ell_\perp^2 = \lambda \pi \ell_\perp^2,$ then

\begin{equation}
{d/dY \rho(\ell_\perp,Y)\over \Delta \ell_\perp^2 dxG/d^2\ell_\perp} = {(2\pi)^2\over \lambda}\ (\alpha_P-1)
\end{equation}

\noindent being independent of $Y$ and $\ell_\perp$ indicates that source gluons having $k_\perp \approx \ell_\perp$ contribute  essentially independently and incoherently to the charge correlator.

Thus the picture of the distribution of color charges in a BFKL wavefunction is quite simple.  Gluons having momentum $\ell_\perp$ appear as completely shielded by other gluons having a similar momentum when viewed on a  scale $\Delta x_\perp > 1/\ell_\perp$ while they appear as completely random charges when viewed on a scale $\Delta x_\perp < 1/\ell_\perp$ with the transition between the two pictures occuring at $\Delta x_\perp \approx 1/\ell_\perp.$ We shall see that this picture is preserved  in the saturation region where BFKL evolution no longer applies.

\section{The light-cone wavefunction in the saturation regime}

To deal with the saturation regime we use the Kovchegov equation\cite{Phy}.  (The Balitsky equation\cite{Bal} would give the same results.)  We take the equation in the form, with $x_\perp = x_{1\perp}-x_{2\perp},$

\begin{displaymath}
{dS(x_\perp, b, Y)\over dY} =  {\alpha C_F\over \pi^2} \int{d^2z_\perp x_\perp^2\over (x_{1\perp}-z_\perp)^2(z_\perp-x_{2\perp})^2}\cdot
\end{displaymath}
\begin{equation}
\cdot \left[S(x_{1\perp}-z_\perp, b, Y) S(z_\perp - x_{2\perp}, b, Y) - S(x_\perp, b, Y)\right],
\end{equation}

\noindent as illustrated in Fig.3.  We suppose the vertices in  the upper parts of the graphs are taken in the large $N_c$ limit, though of course we cannot assume a large $N_c$ limit for all the graphs as unitarity effects would be lost.  We write $C_F$ rather than $N_c/2$ in the prefactor in (22) because that is exact for the virtual correction which will dominate our result.  The factorized form of the two S's on the righthand side of (22), which was originally motivated by Kovchegov by choosing scattering on a large nucleus, cannot be expected to be exact, but this will not be important for the argument which follows.

\begin{center}
\begin{figure}
\epsfbox[0 0 399 79]{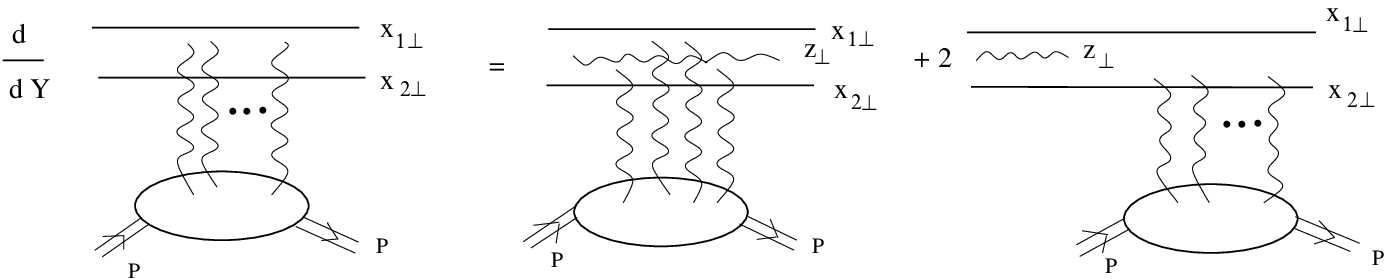}
\caption{}
\end{figure}
\end{center}

If $Q_s^2(Y) x_\perp^2 \gg 1$ the quadratic term in (22) is small when $(x_{1\perp}-z_\perp)^2Q_s^2$ and $(x_{2\perp}-z_\perp)^2 Q_s^2$ are both greater than 1.  Thus the dominant contribution to the righthand side of (22) comes from the second term in the regions

\begin{equation}
1/Q_s^2 \ll (x_{1\perp}-z_\perp)^2 \ll x_\perp^2
\end{equation}

\noindent and

\begin{equation}
1/Q_s^2 \ll (x_{2\perp}-z_\perp)^2 \ll x_\perp^2
\end{equation}

\noindent giving 

\begin{equation}
{dS(x_\perp, b, Y)\over dY} = - {2\alpha C_F\over \pi} \ln[Q_s^2 x_\perp^2] S(x_\perp, b, Y).
\end{equation}

\noindent Using

\begin{equation}
\ln[Q_s^2(Y)/Q_s^2(Y_0)] \approx c {\alpha N_c\over \pi} (Y-Y_0)
\end{equation}

\noindent with \cite{Mue,Tri,ura}

\begin{equation}
c = {2\chi(\lambda_0)\over 1-\lambda_0}
\end{equation}

\noindent where $\chi(\lambda)$ is the usual eigenvalue function appearing in  BFKL evolution, and $\lambda_0$ is determined by ${\chi^\prime(\lambda_0)\over \chi(\lambda_0)} = - {1\over 1-\lambda_0}.$  Thus\cite{vin}

\begin{equation}
S(x_\perp, b, Y) = \exp\left\{{-C_F\over c N_c} \ln^2[Q_s^2(b,Y) x_\perp^2]\right\} S\bigg(x_\perp, b, Y(x_\perp)\bigg),
\end{equation}

\noindent where $Q_s^2(Y(x_\perp)) = 1/x_\perp^2.$

Moving again to a frame where the gluon at $z_\perp$ in (22) and in Fig.3 is in the wavefunction of the hadron on which the dipole scatters, it is natural to identify the exponent in (28) as the scattering, due to two gluon exchange, of the dipole on the saturated hadron.  This gives

\begin{equation}
{C_F\over c N_c}\ \ell n^2[Q_s^2(Y,b) x_\perp^2] = 2\cdot {g^2\over 2 N_c} \int_{1/x_\perp^2}^{Q_s^2} {d^2\ell_\perp\over [\ell_\perp^2]^2}\ {\rho(\ell_\perp, b,Y)\over 4\pi^2}
\end{equation}

\noindent as illustrated in Fig.4 while Eq.29 gives

\begin{center}
\begin{figure}
\epsfbox[0 0 209 74]{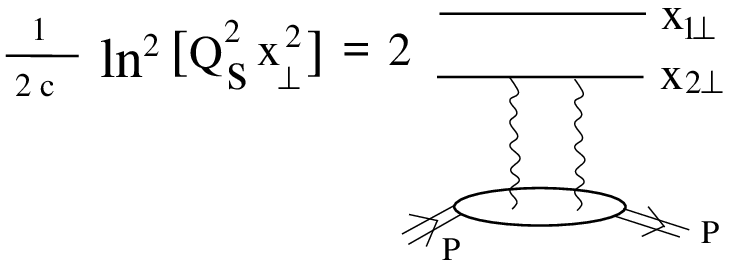}
\caption{}
\end{figure}
\end{center}

\begin{equation}
\rho(\ell_\perp, b, Y) = {2 C_F\over c \alpha} \ell_\perp^2 \ell n[Q_s^2/\ell_\perp^2],
\end{equation}

\noindent where $c$ is given in (27). Now (16) is no longer exactly valid, however, as we shall see in Sec.6, it does continue to have approximate validity.  Using (16) with an undetermined constant  $k$  gives

\begin{equation}
{d xG\over d^2bd^2\ell_\perp} = k {N_c^2-1\over 4\pi^3\alpha N_c}\ {1-\lambda_0\over 2\chi(\lambda_0)} \ell n Q_s^2/\ell_\perp^2.
\end{equation}

\noindent This result is exactly the same as that found in Ref.9 and except for the way the constants are written that of Ref.10 and of the  original McLerran-Venugopalan model, (12).  We believe  it to be a very general result.

\section{Saturation with a running coupling}

In this section we shall generalize the results of Sec.4 to the case of QCD with a running coupling. We begin with the generalization of (25) to the running coupling case which now reads

\begin{equation}
{dS(x_\perp, b, Y)\over dY} = - {2 C_F\over \pi b} \ell n\left[{\ell n Q_s^2/\Lambda^2\over -\ell n x_\perp^2\Lambda^2}\right] S(x_\perp, b, Y).
\end{equation}

\noindent The analogue of (26) is\cite{Tri,ura}

\begin{equation}
\ell n\left[{Q_s^2(Y,b)\over \Lambda^2}\right] = {\sqrt{{2N_c\over \pi b} c Y}}\ + O(Y^{1/6})
\end{equation}

\noindent where  $c$  is exactly as given by (27).  Combining (33) with (32) yields

$$
S(x_\perp, b, Y) = \exp\left\{{-C_F\over 2 cN_c}\bigg [2\ln^2 Q_s^2/\Lambda^2 \ln\left({\ln Q_s^2/\Lambda^2\over -\ln x_\perp^2\Lambda^2}\right)-\ln Q_s^2x_\perp^2\ln {Q_s^2\over x_\perp^2\Lambda^4}\bigg]\right\}\cdot 
$$
\begin{equation}
\cdot S\left(x_\perp, b, Y(x_\perp)\right)
\end{equation}

\noindent which now replaces (28).

Using the exponent in (34) to identify $\rho,$ exactly as was done in (29), gives

\begin{equation}
\rho(\ell_\perp, b, Y) = {2C_F\over c} <{1\over \alpha}> \ell_\perp^2 \ln[Q_s^2/\ell_\perp^2]
\end{equation}

\noindent with

\begin{equation}
<{1\over \alpha}> = {1\over 2} [{1\over \alpha(Q_s^2)} + {1\over \alpha(\ell_\perp^2)}].
\end{equation}

\noindent Finally using (16), but again with an undetermined constant, $k,$

\begin{equation}
{d xG\over d^2b d^2\ell_\perp} = k {N_c^2-1\over 4\pi^3N_c} <{1\over \alpha}> {1-\lambda_0\over 2\chi(\lambda_0)} \ln Q_s^2/\ell_\perp^2
\end{equation}

\noindent as in (31) except for the replacement ${1\over \alpha} \rightarrow <{1\over \alpha}>.$

Thus running coupling effects do not essentially change (30) and (31) which now take the forms (34) and (36), respectively.

\section{A simple physical picture}

In this section we present a simple physical picture of $\rho(\ell_\perp, b, Y)$ and of ${d x G\over d^2b d^2\ell_\perp}$ in terms of which gluons in the saturated wavefunction can be viewed as determining $\rho$ and ${d x G\over d^2b d^2\ell_\perp}.$  The picture is complementary to the picture given in Ref.9 where ${d x G\over d^2\ell_\perp d^2b}$ was given a simple interpretation in terms of scattering of a gluon dipole on a hadron.  In Fig.5 we exhibit the region of gluons in saturation using the fixed coupling boundary given in (26).  We shall carry out the discussion in a fixed coupling context, however, the picture is identical for the running coupling case.  In Fig.5 the higher momentum gluons are at the top of the figure, with small values of $y,$ while softer gluons are at the bottom of the figure with large values of $y.$  In Fig.6 we plot the $x_-$ extent of the gluons in the wavefunction using the Kovchegov gauge\cite{Kov} where final state rescatterings of produced gluons are absent.  The $x_-$ boundary of the gluons in Fig.6 is\cite{Ian}$x_-P_+\approx  - e^y$ where $P_+$ is the momentum of the hadron.

\begin{center}
\begin{figure}
\epsfbox[0 0 199 158]{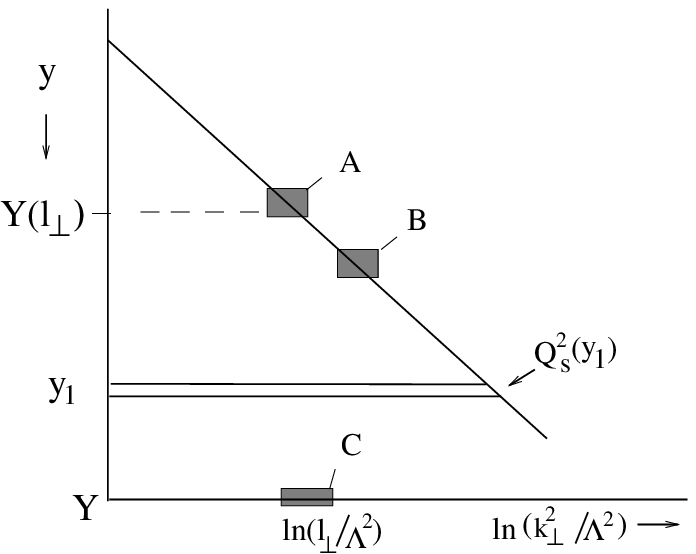}
\caption{}
\end{figure}
\end{center}

We begin our discussion by considering $d\rho/dY$ which, from (30) and (26), is

\begin{equation}
{d\rho(\ell_\perp, b, Y)\over dY} = {N_c^2-1\over \pi} \ell_\perp^2.
\end{equation}

\noindent The gluons contributing to (38) come directly from parent gluons located in the region \  A\   of Fig.5.  Region A is at the edge of the saturation boundary and is of size $\Delta \ell_\perp \sim \ell_\perp, \Delta y \sim 1/(\alpha N_c).$  Thus there are on the order of ${(N_c^2-1)\ell_\perp^2\over (\alpha N_c)^2}$ gluons per unit area in \ A.\ These gluons emit on the order of ${(N_c^2-1)\ell_\perp^2\over (\alpha N_c)^2}\cdot \alpha N_c$ gluons per unit area and per unit rapidity directly into the region $y \approx Y, k_\perp \approx \ell_\perp.$  On a scale $\ell_\perp$ these gluons are essentially uncorrelated in color leading to

\begin{equation}
{d\over dY} \rho(\ell_\perp, b, Y) \sim (N_c^2-1) \ell_\perp^2
\end{equation}
\noindent agreeing with (38).  Two natural questions arise.  (i)  Why not include emissions from other areas, such as region B of Fig.5?  Why are there no secondary emissions starting from A?  That is a  source lying in\  A\  could emit a gluon at transverse momentum on the order of $\ell_\perp$ and at rapidity  $y_1,$ lying between $Y(\ell_\perp)$ and $Y,$ and this gluon could then emit a secondary gluon having $\ell_\perp$ and $Y.$  The answer to (i) is that gluons coming from region  B  are highly shielded on a scale $\ell_\perp$ and contribute little to charge correlations, although, as we shall see shortly they do contribute to the total {\underline {number}} of gluons at $\ell_\perp$ and $Y.$  The answer to (ii) is that gluons emitted from $y_1$ will typically have transverse momentum on the order of $Q_s(Y_1).$  The probability that a gluon be emitted from $\ell_\perp,y_1$ to $\ell_+, Y$ is proportional to $\ell_\perp^2/Q_s^2(y_1) << 1.$  Thus the gluons that contribute to ${d\rho(\ell_\perp, b, Y)\over dY}$ are themselves gluons having transverse momentum on the order of $\ell_\perp$ and which come by direct emission from the region near $\ell_\perp, Y(\ell_\perp)$ where $Q_s[Y(\ell_\perp)] = \ell_\perp.$

Next we turn to ${d x G\over d^2\ell_\perp d^2b}$ evaluated at $Y.$  Now gluons coming from sources having rapidities between $Y(\ell_\perp)$ and $Y$ are important.  Focus on sources at rapidity $y_1.$  We can expect a contribution of size

\begin{equation}
{d x G\over d^2b} \sim {Q_s^2(y_1)\over \alpha} \cdot \alpha \cdot {d^2\ell_\perp\over Q_s^2(y_1)} \cdot dy_1
\end{equation}

\noindent coming from source in a range $dy_1.$  Of the four factors on the right-hand side of (40) the first, $Q_s^2/\alpha,$ is the number of gluons per unit area and  per unit rapidity at $y_1.$  The second factor, $\alpha,$ is the strength of emission of gluons from $y_1.$  $d^2\ell_\perp/Q_s^2$  is the probability that the emitted gluon end  up in the momentum range $d^2\ell_\perp.$  (Recall that in the Kovchegov light-cone gauge a gluon emitted from $y_1$ is gauge rotated, locally, by all gluons having larger  values of $x_-$\cite{ler}, that is essentially by all gluons having rapidities greater than $y_1$ as is clear from Fig.6.  These gauge rotations make the typical transverse momentum of gluons emitted from sources at $y_1$ on the order of $Q_s(y_1).) dy_1$ is the range of sources in rapididty.  Integrating $dy_1$ between $Y(\ell_\perp)$ and $Y$ gives\cite{Mue,Ler}

\begin{center}
\begin{figure}
\epsfbox[0 0 242 191]{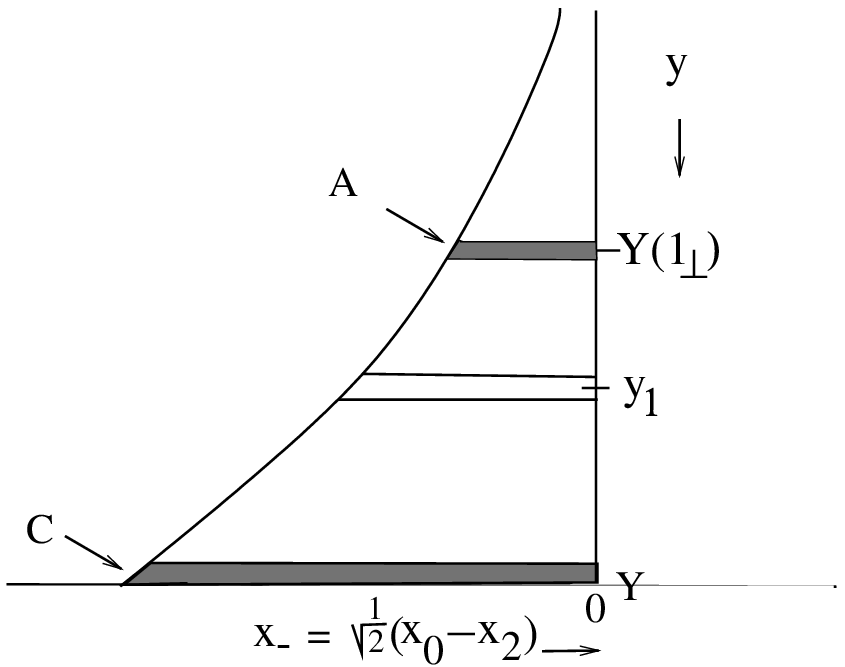}
\caption{}
\end{figure}
\end{center}

\begin{equation}
{d x G\over d^2\ell_\perp d^2b} \sim [Y-Y(\ell_\perp)] = {\pi\over \alpha N_c}\ {1-\lambda_0\over 2\chi(\lambda_0)} \ln {Q_s^2(y)\over \ell_\perp^2}
\end{equation}

\noindent which contains the essential factors in (31) and (35).  It is important to note that the gluons contributing
 to (39) are highly correlated in color.  {\underline {If}} they were randomly distributed in color they would lead to a contribution to ${d\rho(\ell_\perp, b, Y)\over dY}$ of size $\alpha[Y-Y(\ell_\perp)]\sim \ln[Q_s^2(Y)/\ell_\perp^2]$ which is too large.  Because the sources from which most of the gluons in (41) come from are strongly shielded on scale $\ell_\perp$ the gluons in (41) should themselves contribute little to $\rho(\ell_\perp, b, Y).$  Finally, gluon sources at $y_1$ having momentum $k_\perp \ll Q_s(y_1)$ are too few in number to contribute to (40) while gluon sources having $k_\perp \gg Q_s(y_1)$ are too strongly shielded to be important.

It is interesting to carry out one last calculation which helps explain why (28) and (34) have all $Y-$dependence occurring in the exponent.  To that end we turn to an interpretation of (25). In the Kovchegov picture the fact that ${dS\over dY}$ is proportional to $S$ comes about because the virtual correction shown in Fig.3 occurs long before the actual scattering, the term shown in the figure, or long after the scattering.  The exponential result then comes about because the virtual gluon corrections are ordered in time, according to the longitudinal momentun of the gluons,  thus naturally leading to an exponential result.  How does this occur in a picture where the $Y-$dependence is put into the target rather than in the dipole?  When the rapidity of the target is increased by $dY,$ there is an increase in the charge correlator by $d\rho(\ell_\perp,b,Y)$ and this increased charge is located at ``large'' negative values of $x_-$ as shown in the shaded part of Fig.7 where the top part of that figure shows the distribution of charge as a function of $Y,$ much as in Fig.6.  Also shown is  a scattering off that additional charge by the dipole which reaches the charge in $d\rho$ before it reaches the charge located at higher rapidities.  There must be at least two gluons exchanged because the charges located at different rapidities are too weakly correlated to give a non-zero result and there will be no more than two gluons going from $dY$ to the dipole since the coupling to the charge in $dY$ is not very large unless $dY$ is quite big. More precisely we can rewrite (25) as

\begin{center}
\begin{figure}
\epsfbox[0 0 207 162]{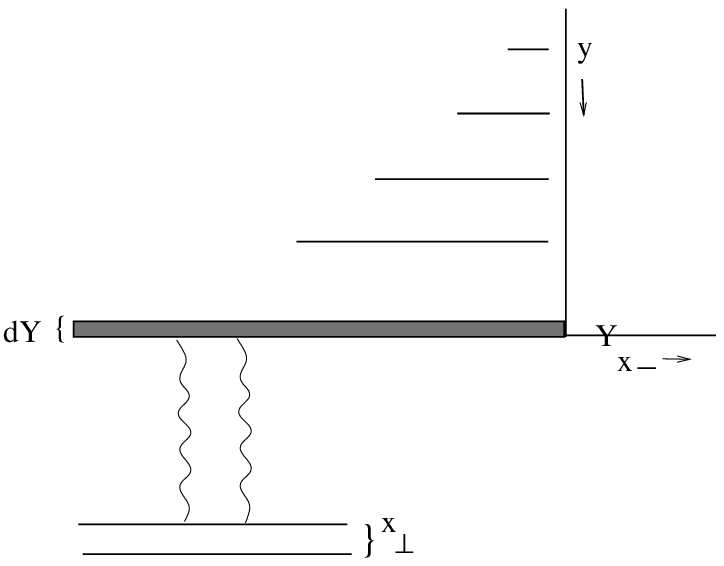}
\caption{}
\end{figure}
\end{center}

\begin{equation}
{dS\over dY} (x_\perp,b, Y) = - \int {d\rho(\ell_\perp, b,Y)\over dY}\ {d^2\ell_\perp\over 4\pi^2[\ell_\perp^2]^2}\cdot  2\cdot  {g^2\over 2N_c} S(x_\perp, b, Y).
\end{equation}

\noindent The factors on the righthand side of (42) are: $g^2/2N_c$ is the coupling of two gluons to one of the quarks in the dipole; the 2 counts the number of quarks in a dipole while ${1\over 4\pi^2}\ {d^2\ell_\perp\over [\ell_\perp^2]^2}$ gives the phase space and propagators.  $S(x_\perp, b, Y)$ then corresponds to the later, in time, scatterings with the rest of the charge in the hadron's wavefunction.  The integral in (40) goes from $\ell_\perp = 1/x_\perp$ to $\ell_\perp = Q_s(Y).$  Using (38) then leads to (25).

\end{document}